\documentclass[12pt]{article}

\usepackage[pdftex]{graphicx}
\usepackage{scicite}

\usepackage{times}

\newenvironment{sciabstract}{%
\begin{quote} \bf}
{\end{quote}}

\newcounter{lastnote}

\title{Reconfigurable Application-Specific Photonic Integrated Circuit for solving Partial Differential Equations}

\author
{Chen Shen$^{1}$, Nicola Peserico$^{1}$, Jiawei Meng$^{1}$, Xiaoxuan Ma$^{2}$, \\Behrouz Movahhed Nouri$^{2}$, Cosmin-Constantin Popescu$^{3}$, \\ Juejun Hu$^{3}$,Tarek A. El-Ghazawi$^{1}$, Hamed Dalir$^{2}$, Volker J. Sorger$^{1,2,\ast}$\\
\\
\normalsize{$^{1}$The Department of Electrical and Computer Engineering, George Washington University,}\\
\normalsize{800 22nd St NW, Washington, DC 20052, USA}\\
\normalsize{$^{2}$Optelligence LLC, 10703 Marlboro Pike, Upper Marlboro, MD 20772, USA}\\
\normalsize{$^{3}$Department of Material Science and Engineering, Massachusetts Institute of Technology,}\\ \normalsize{Cambridge, MA, USA}\\
\\
\normalsize{$^\ast$To whom correspondence should be addressed; E-mail:  sorger\{@gwu.edu;@optelligence.co\}}
}

\date{}

\begin{document} 


\baselineskip24pt


\maketitle


\begin{sciabstract}
 Solving mathematical equations faster and more efficiently has been a Holy Grail for centuries for scientists and engineers across all disciplines. While electronic digital circuits have revolutionized equation solving in recent decades, it has become apparent that performance gains from brute-force approaches of compute-solvers are quickly saturating over time. Instead, paradigms that leverage the universes’ natural tendency to minimize a system's free energy, such as annealers or Ising Machines, are being sought after due to favorable complexity scaling. Here we introduce a programmable analog solver leveraging the mathematical formal equivalence between Maxwell's equations and photonic circuitry. It features a mesh network of nanophotonic beams to find solutions to partial differential equations. As an example, we designed, fabricated, and demonstrated a novel application-specific photonic integrated circuit comprised of electro-optically reconfigurable nodes, and experimentally validated 90\% accuracy with respect to a commercial solver. Finally, we tested this photonic integrated chip performance by simulating thermal diffusion on a spacecraft’s heat shield during re-entry to a planet's atmosphere. The programmable light-circuitry presented herein offers a facile route for solving complex problems and thus will have profound potential applications across many scientific and engineering fields.
\end{sciabstract}


\section*{Introduction}
\begin{figure}[h!]
    \includegraphics[width=\textwidth]{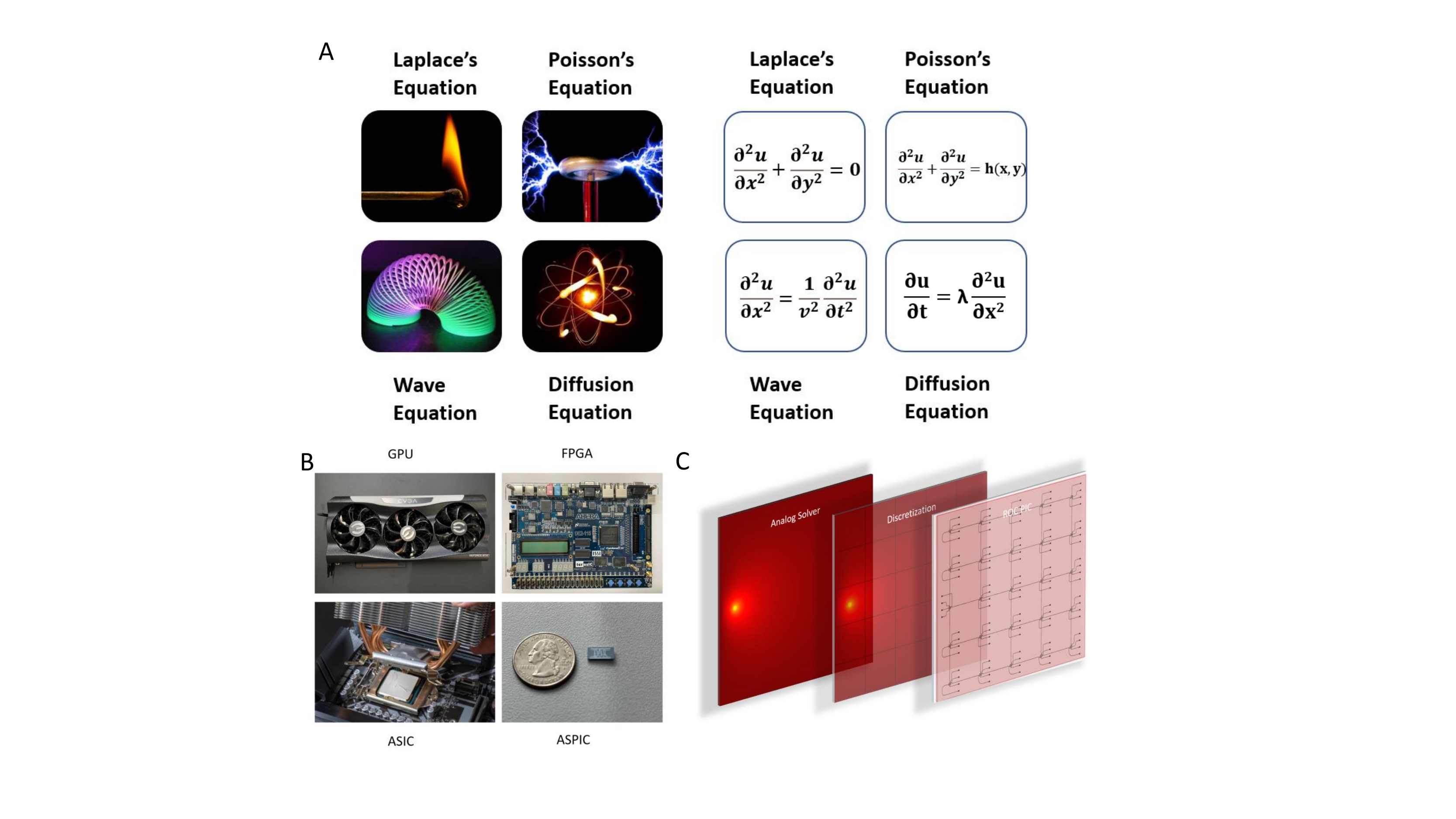}
    \centering
    \caption{(A).Different classification of partial differential equations. (B). Device sizes comparison of GPU, FPGA, ASIC, and ASPIC. (C). Mapping from analytic solution of PDEs to the photonic solver.}
    \label{fig:fig1}
\end{figure}

In the mathematical field, equation solving has been a key action to progress in the evolution of the human species. For most of the human history, this task was performed by hand, involving long paper-written equations and solutions to be computed. First digital electronics and then computers have turn equation solving task into a simpler paradigm, exploiting digital computation and algorithms to achieve the searched results. Beside the close-form solutions algorithm, the main algorithm to solve equations still relies on brute-force search, where a solution is searched by exhausting the entire solution space.
\\
However, it is straightforward that this approach brings many limitation, such as energy consumption as well as the huge amount of time needed, if the solution space is wide. Nowadays, a quantum annealer like D-wave or Coherent Ising Machine can solve NP-hard problems in polynomial time\cite{hamerly2019experimental,inagaki2016coherent} by relying on other mechanism, inherited by the physics of the devices themselves, although the size of both two solvers is bulky in reference to an integrated chip. Similar to many unsolved problems in Computing Science that may be solved by the P vs NP problem and Rieman Hypothesis, here we would like to introduce an analog solver to solve a specific class of Partial Differential Equations (PDEs) in small size and a time-of-flight of photon's speed.\\ \\
First of all, PDEs are a widely known way to mathematically describe various amounts of problems, physical laws, and even economic behaviours\cite{renardy2006introduction}. There is a long history of human beings solving PDEs since their definition was introduced in the 18th century by Euler, d’Alembert, Lagrange and Laplace\cite{widder2015laplace}. There are several examples where PDEs are used in models to precisely describe behaviors and laws, and their rapid solution is therefore of significant practical importance. In a few instances, models that use PDEs could have precise closed-form solutions easily computed using mathematical formulas (e.g. propagation of an electromagnetic plane wave in a uniform medium). In general for most problems, however, close solutions cannot be found, leaving numerical solutions as the only way to achieve approximate results (e.g. for many models in fluid dynamics models that rely on Navier–Stokes equations\cite{temam2001navier}, as shown in figure \ref{fig:fig1}A).   In fact, mathematicians, scientists and engineers have spent innumerable time and effort optimizing the solving mechanism. \\
Currently, automatic digital computation has brought a powerful tool to solve numerically PDEs, allowing recursive interaction to find the approximate results\cite{braun1995numerical}. The rise of CMOS as a way to fabricate computing units (such as CPU), where billions of transistors can be set to perform millions of mathematical operations per second, has paved the way for solving PDEs models making use of commercial software\cite{yentis1996vlsi}. The use of digital solvers has several advantages, in particular accessibility to interactive recursive approximation algorithms capable of iteratively improving solution accuracy. However, this approach comes with a major drawback, namely the excessive time required when approaching complex PDEs. As a consequence, large data centers have been used to process complex models, for example in multi-physics problems. To overcome such limitations, analog solvers have been studied\cite{liebmann1950solution}, which mimic complex physical models and obtain solutions in a fraction of time required by digital solvers. \\ \\
Here, we present a tunable-weights Application-Specific Photonic Integrated Circuit (ASPIC) which offers a state-of-art option for solving PDEs. By using photonics we can take advantage of the extremely large bandwidth as well as the extremely low latency and power consumption to mimic the PDE models and obtain approximate results, with accuracy that reaches up to 93\% as benchmarked by commercial digital solvers. Moreover, our proposed solution consumes a negligible amount of energy compared to state-of-the-art electronic CPUs or GPUs (Figure \ref{fig:fig1}B). \\ \\
In this paper, we first introduce the PDE model we address, and the related ASPIC as a PDE solver using photonic integrated circuits, showing how the ASPIC can map the actual physical problem, as in figure \ref{fig:fig1}C. We follow with characterization of our photonic solution using the Kirchhoff's Photonic Nodes (KPNs), and a photonic network implemented with KPNs. The last part will show passive and active ASPIC networks that mimic the diffusion of heat into homogeneous (passive network) and non-homogeneous (active network) media, highlighting the high accuracy attainable by our solution. The last part is dedicated to discussion of future improvements.

\section{Result}
\begin{figure*}[h]
 \centering
 \includegraphics[width=\textwidth]{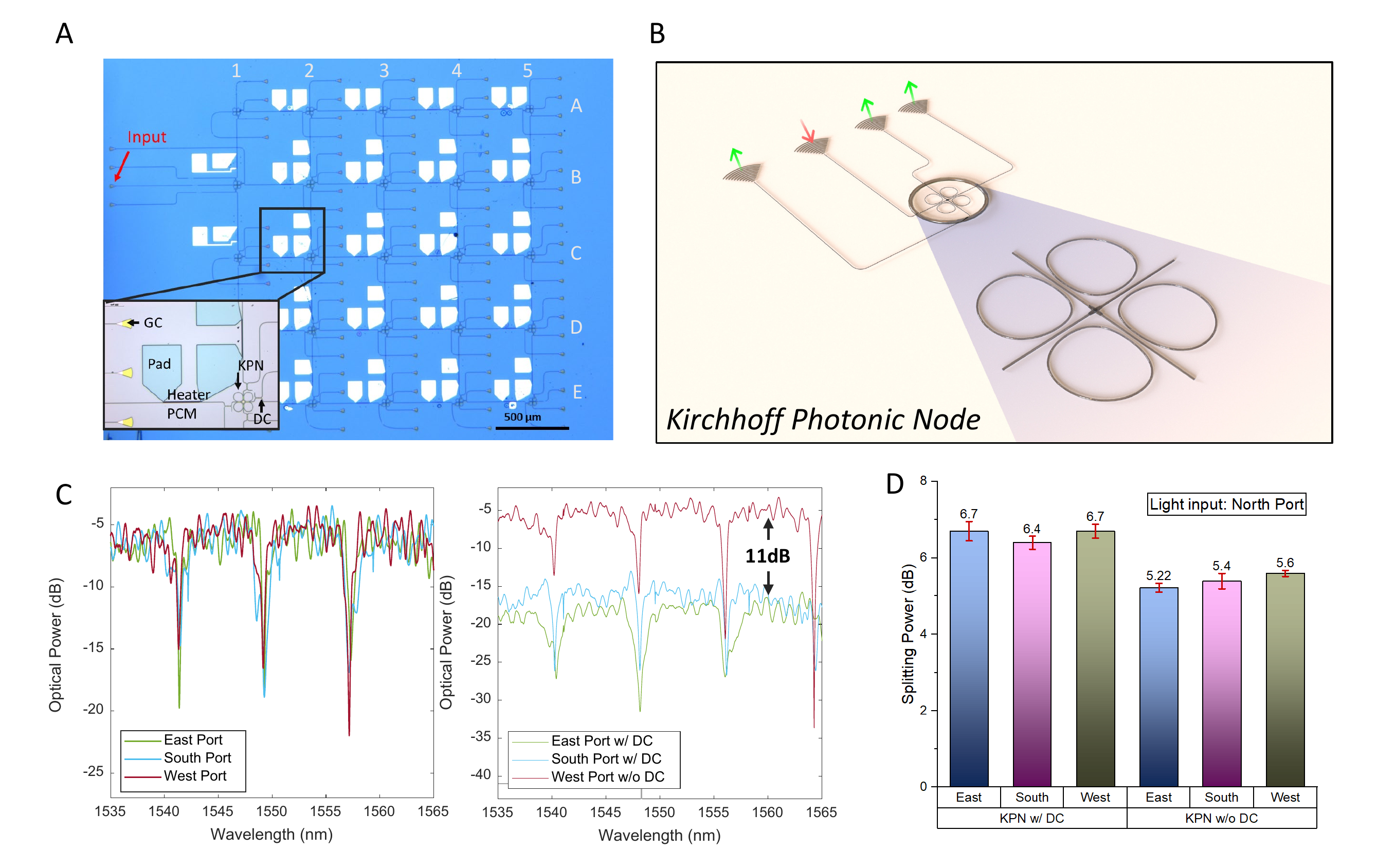}
 \caption{Passive ASPIC (A).The ASPIC image taken by Leica DMi8 Inverted Fluorescent Microscope. (B).Schematic of a Kirchhoff Photonic Node. To measure the splitting ratio, we inject the light into the grating coupler and transmit it through one waveguide. The other three grating couplers connecting to three ports will couple the light vertically out for readout. (C). Two spectra show that the KPN's port connected to a direction coupler will divide 10\% of the light for measurement as well as the performance stability of the KPN. (D). These columns with red error bars in this chart indicate that the splitting ratio will not change with the directional coupler implementation.}
 \label{fig:fig2}
\end{figure*}

\paragraph{ASPIC as a PDE Solver}
$ $ \\
We aim to solve the heat distribution over a homogeneous medium for our demonstration by mapping this problem to our ASPIC. This PDE-based model can be configured with different parameters and boundary conditions, such as different thermal conductivity of the materials or the heating flux at the boundary of our model.\\ 
While electronics, in particular resistors networks, can be a good analog solver for PDEs, given the intrinsic behavior of electromagnetic waves on lumped elements, photonics finds itself lacking these possibilities, as the wave propagation does not affect the previous elements in the first place (i.e. a modulator after one branch of a 1:2 splitter does not affect the other branch). To break this limitation, we introduced a photonic matrix network whose crossing nodes are formed by a four-port Kirchhoff’s Photonic Node (KPN) \cite{sun2021induced}\cite{shen2021application}, shown in figure \ref{fig:fig2}B. This four-port component has the property to split the light from one port to the other in a quasi-even fashion. This node element mimics the Kirchoff’s current law in optics, while inheriting all the advantages that optics brings, such as high speed, low latency, and robustness to electromagnetic interference. \\ 
\paragraph{KPN using Silicon Photonics}
$ $ \\
A series of KPNs and a full 5x5 matrix were realized on a Silicon Photonics passive platform\cite{bogaerts2018silicon}. We chose 500 nm waveguide width, using periodic grating couplers\cite{wang2013universal} to couple the light from the light source to the chip, as well as from the chip to the photodetectors, as shown in figure \ref{fig:fig2}B. Similar to the electrical counterpart, we need to extract the total power out from all the nodes present in the network. To do so, while reducing the losses that a direct detection would introduce, we place a 90\%:10\% Directional Coupler (DC) at the output of all the node's ports, directing the smaller amount of light to a grating coupler. By doing so, we can compute the power in the node by summing the power coming from all the ports, while the larger part of the light signal will continue toward the next node. The design of the DC was performed following Lu et al. method\cite{lu2015broadband}. We fabricate several different KPNs having different DC configurations that reflect the ones present in the main matrix. For example, a KPN placed on the corner of the matrix will have just 2 DC, as the ports that are not facing another node do not require a splitter before the grating coupler. Figures \ref{fig:fig2}C show the spectra of two KPNs, where light is coupled from the north port. The effect of the DCs are clearly visible in the second spectrum, where East and South ports have a 10 dB reduction compared to the West one. To notice that, as we consider the DC presence or absence among a series of 9 different configurations of KPN, we statistically see a small difference, as shown in figure \ref{fig:fig2}D. Moreover, the errors due to fabrication variability are reasonably contained.\\ 
The general spectra of the KPN shown in the same figure \ref{fig:fig2}C show a response resembling that of a ring resonator coming from the structure of the used KPN. In particular, KPN responses have a Free Spectrum Range of 8 nm (998 GHz) at 1550 nm, an average bandwidth of 1.26 nm, resulting in a Q-factor of 1000. This component can reach a high Extinction Ratio of more than 15 dB. \\
\paragraph{Silicon Photonics ASPIC}
$ $ \\
Using the defined KPN, we design and fabricate a 5x5 passive silicon photonic matrix, as shown in figure \ref{fig:fig2}A. We placed each node to be 500 $\mu$m apart, linking them with a straight silicon waveguide, having a 2.1 dB/cm estimated propagation loss. The input of our matrix is placed in node 1-B (using a coordinate system from the figure). In our heating model, this will be the place where the heating source is set.
The matrix we implement is a larger version of the previous one (from \cite{sun2021induced}), allowing a better resolution of the model we solve. A larger matrix is still possible. However, propagation losses, as well as optical power dynamic range, must be taken into account. For example, considering using the commercial laser with a power output of 10 dBm and photodetectors with sensitivity of -70 dBm, we can compute that the largest matrix the approach could afford will be a 6x6 one considering node splitting, as well as propagation and coupling losses. Some strategies can be used to address this limitation, for example, by using Silicon Nitride (SiN) waveguides that permit lower propagation losses, as well as other strategies like CLIPP\cite{grillanda2014non} to detect the output power from a single node that does not require a DC. \\

\paragraph{Heat Model Benchmark}
$ $ \\
In this problem, we are addressing a subset of PDEs describing heat distribution within a uniform medium. In particular, we apply these heating equations to a 2D surface, having a heat source located at the input of our 5x5 matrix and heat flux at the boundary of the surface.
We obtain our benchmark model result from COMSOL multiphysics, a known commercial and solid numerical iterative solver. By using the COMSOL solution, we can benchmark the accuracy of the solutions of a Laplace homogeneous PDE obtained by our ASPIC. The COMSOL model is formed by a 25 m$^2$ square film made of a fixed material (whose heat conductivity is set). A fixed temperature point is applied at the left top of the geometry, applying a constant temperature (393 K) as a heat source. Simultaneously all the remaining edges are set to have a heat flux toward the outside, which is set at 293 K to represent the heat sink of this model. We set the computation domain to be divided into 25 smaller squares, represented by our 5x5 model and computed the average temperatures of all these squares. This model can simulate the actual thermal transfer profile across the domain under the Dirichlet boundary conditions (BC).

Besides our experimental ASPIC, we built a one-on-one 5x5 optical mesh on the Lumerical Interconnect software to simulate the ASPIC and imitate the heat transfer mesh structure. At the crossing nodes, we used the scattering matrix parameter (S-parameter) of the KPN, computed to reproduce its optical characteristics. Same as our Silicon Photonic ASPIC, we extract the power from each node by placing a DC in all the ports of the KPN. Moreover, all the nodes are connected by a waveguide model, whose propagation loss is set to the experimental value and tuned to match the heat conductivity of our model. \\

\begin{figure*}[h!]
 \centering
 \includegraphics[width=\textwidth]{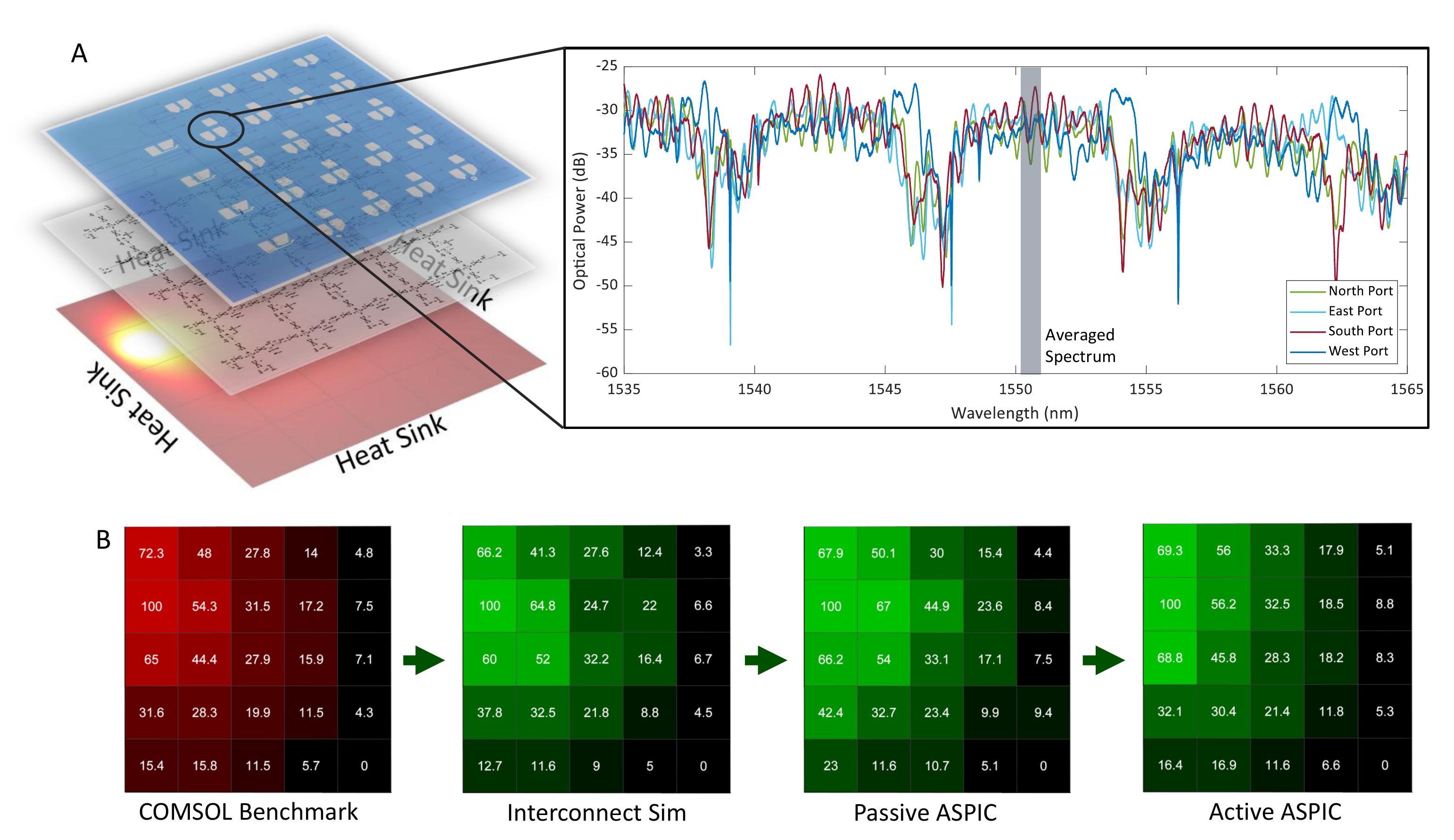}
 \caption{ASPIC PDE solver (A). The stacking diagrams are the correlation of different solver's power map from COMSOL solver to Interconnect solver and ASPIC solver. The analysis figure along the diagrams shows the power readout spectrum of one KPN in the ASPIC, indicating that the power splitting of four ports is analogous. (B). These power maps demonstrate the thermal distribution in COMSOL solution and optical power distribution in Interconnect and ASPIC solver. (C). From the Accuracy chart, we can see that both accuracy of Interconnect and passive ASPIC can not reach above 90\%}
 \label{fig:fig3}
\end{figure*}

\paragraph{Passive ASPIC Results}
$ $ \\
To measure our passive ASPIC, we inject a 10 $mW$ tunable continuous-wave laser, centered at 1550 $nm$, through the grating coupler at the input node. We then collect all the power output from all the ports of all the nodes to compute the actual power for our model. A spectrum of the four outputs from one KPN is shown in figure \ref{fig:fig3}A. The spectra are similar to the one presented in the previous section, with the same FSR. However, in this case, we can notice that the notches do not have the same shape or bandwidth as those of a single KPN. Since this node is inside a matrix, its spectrum is the result of all the incoming light beams from adjacent nodes that have slightly de-tuned microring notches, that results in a more complex line shape. Moreover, the non-optimized grating couplers produce a Fabry-Perot ripple that we eliminate by averaging over one period, similar to having a large bandwidth laser at the input. \\
After collecting the data from the measurement and from the simulations, considering that those quantities are incoherent in terms of measured units, we opted for a normalization scheme (0-100). The data are shown in figure \ref{fig:fig3}B, as a heat map of COMSOL, Interconnect simulation, and experimental data from the passive ASPIC.\\ \\
From these results, we can define the accuracy as:
\begin{equation}
A=1-\sum_{i=1}^n \frac{N_{i,c}-N_{i,a}}{_{Ni,c}}, \quad
\end{equation}
where $A$ is the total accuracy of the solution, $n$ is the number of the nodes, $N_{i,c}$ and $N_{i,a}$ are respectively the node values of COMSOL solution and solver solution (simulated or measured), normalized to a 0-100\% range. With our model, we obtain a 89.5\% accuracy for the Interconnect solution, and a 86.4\% accuracy for the experimental ASPIC (Figure \ref{fig:fig3}C). \\
The obtained accuracy is slightly below the minimum standard required for a solver (90\% or above), and it is mainly due to the mismatch of optical properties over the matrix, as well as the passive structure that does not allow tuning the losses between nodes to match the physical heat properties of the material in the model. However, the ASPIC solver can compute the model with a latency of hundreds of ns, compared to the COMSOL model, which requires several ms, making the optical solution several orders of magnitude faster.

\begin{figure*}[h!]
 \centering
 \includegraphics[width=\textwidth]{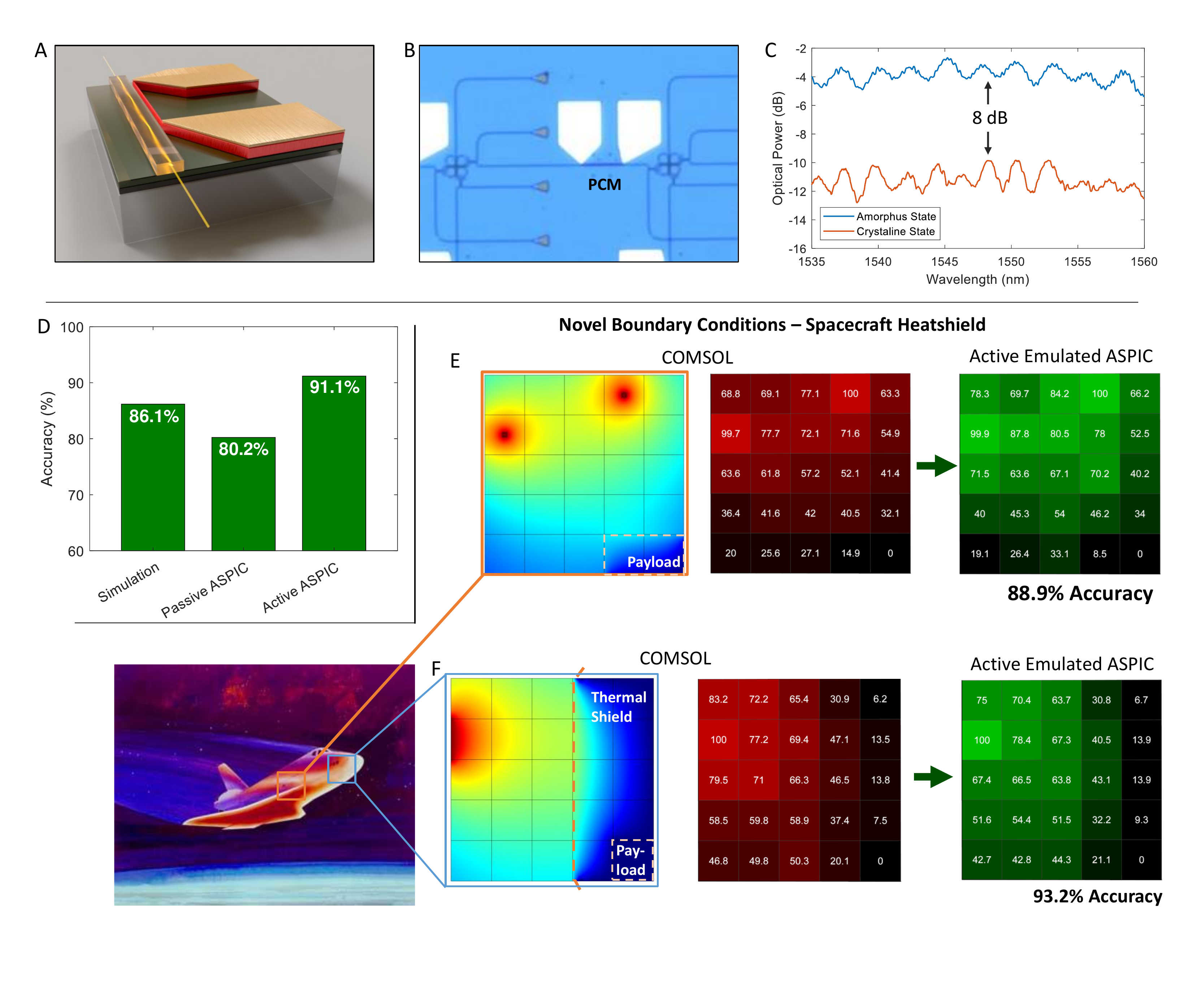}
 \caption{PCM implemented full active ASPIC solver. (A). 3-D schematic diagram of PCM and heater design, patterning using electron-beam lithography and sputter deposition for dual metal layers. (B). The optical image of the heater and contact pad is shown in panel A. (C). This plot shows the spectrum of waveguide power detected with contrast states of the PCM strip employed. By applying a hybrid pulse to the heater, the PCM switched from amorphous state to crystalline state, increasing the extinction ratio by 0.08 $dB/\mu m$.(D). Comparison between the COMSOL heatmap and the power map obtained by the active ASPIC. In this case, the ASPIC can solver PDEs with a stable accuracy of over 91\%. (E)(F). Two heat models for different Spacecraft cases including one or two heat sources and Heatshield. The heatmap and power maps show the modeling of the Thermal Shield that is protecting the Payload under the re-entry phase of a space journey. Even in these demanding cases, the accuracy of our Active ASPIC remains higher than 88.9\%. }
 \label{Fig:fig4}
\end{figure*}

\paragraph{Active ASPIC as PDEs solver using PCM}
$ $ \\
To further improve the accuracy of our ASPIC, a tunable element has to be placed between each node pairs to increase the losses and by so matching any material properties and boundary conditions. Among all the components we can use for this scope, such as Mach-Zendher interferometers, Variable-Optical Attenuator, and so on, we chose a non-volatile tunable element based on an integrated Phase Change Material (PCM), whose optical properties can be modulated with its crystalline structure. This choice enables ultra-compact integration, reduced system complexity, and high energy efficiency taking advantage of the non-volatile nature of PCM. Among the many PCM materials, we select GSSe ($Ge_2Sb_2Se_5$) since it provides an excellent Extinction Ratio (ER) between its amorphous and crystalline states, while having an almost negligible Insertion Loss at the C-band.
\paragraph{Phase Change Materials for Tunable Component }
$ $ \\
The integration of the GSSe has been done by depositing a thin strip over the waveguide and implementing a heater scheme to provide the Joule heating energy to switch the material between its two states\cite{meng2020multi}. A graphical representation of the implementation is shown in figure \ref{Fig:fig4}A. We place this component between each KPN node, as shown in figure \ref{Fig:fig4}B, permitting full reconfigurability of the ASPIC. \\

For the fabrication, we deposited a GSSe strip with 40 nm thickness and 100 $\mu$m length on top of the silicon waveguide, covered with a 10 nm $Al_{2}O_{3}$ layer to prevent oxidation of GSSe. Then, a 350 nm thick Tungsten-Titanium (W/Ti) alloy heater with identical length is paralleled, laying 1 $\mu$m away from the waveguide, routing to two 100 $\mu$m $\times$ 100 $\mu$m contact pads. On top of the contact pads and route, we deposit a 200 nm thick Aluminum layer to minimize the resistance. After the fabrication process, we measured each heater, resulting in an 11 $k\Omega$ resistance. \\
A heuristic approach is taken to study the sufficient power required for switching the GSSe's state. In accordance with the measurement results, we observed a hybrid pulse comprised of a 12 $V$ high voltage for 500 $\mu$ s and a 9.6 V low voltage for 2 $ms$ could switch the GSSe strip from the amorphous state to the crystalline state (Figure \ref{Fig:fig4}C). The spectrum plotted indicates that the absorption is 8 dB, which implies a 0.08 dB/$\mu$m absorption coefficient of the GSSe on-chip. Significantly, the imaginary part of the effective refractive index in the amorphous state is $2.18 \times 10^{-5}$, leading to an exceedingly small unit of passive insertion loss \cite{meng2022electrical}.\\

\paragraph{Active ASPIC as PDE solver}
$ $ \\
The tunable PCM elements implemented in our circuit impart full programmability to the ASPIC. Following a heuristic approach of switching different states in single or multiple GSSe strips within the matrix, we monitor the accuracy variation and seek the highest accuracy value. Over the course of time, by selectively switching the GSSe strips from the amorphous state to the crystalline state, we successfully achieved the solution corresponding to the same problem, solved by the Interconnect solver and passive ASPIC solver, with an accuracy of 93.2\% (Figure \ref{Fig:fig4}D), increased by 7.8\% from the full passive ASPIC. \\ \\
To demonstrate the flexibility of our active ASPIC, we propose two more approximate models of heat shields present in all space vehicles\cite{rasky2012perspective}. In both cases, this shield has to protect the whole vehicles against high-temperature environments in different parts of the vehicle body, particularly during the re-entry phase into the atmosphere, when the vehicle gets exposed to temperatures of more than 1000°C\cite{gordon2014space}. In the first case, we emulated a thermal shield with low heat conductivity in dual high temperature sources situation, that has the main goal to shield the bottom right square from the higher temperature source in figure \ref{Fig:fig4}E. After obtaining the COMSOL result as a benchmark, we iteratively switched various PCM strips between different nodes in the photonic circuit following the same heuristic approach. Thus, we can program the ASPIC to adapt the new model with 88.9\% accuracy, as show in the latter heatmaps of figure \ref{Fig:fig4}E. Similarly, the second COMSOL model presents now a high temperature source, and a thermal shield. The first heatmap in figure \ref{Fig:fig4}F shows the actual temperature distribution and the strong thermal insulation provided by the heat shield, and the accuracy of this model is 93.2\%. We can also observe from the power map comparison that the highest error occurs at the first layer in the heat shield. The explanation is that our PCM “modulators” only have a fixed tunable extinction ratio of 8 dB, and so, we can not arbitrarily tune the extinction ratio precisely in various states. However, by means of applying more PCM strips with different lengths deposited on the waveguides, we will be allowed to address this problem and increase the accuracy.

\section{Discussion}

As we showed, it is possible to solve the PDE model using photonic integrated circuits, achieving high accuracy. However, the circuit shows some limitations that must be taken into account.\\ 
First of all, the separation between two adjacent nodes sets the minimum amount of loss (as propagation loss) that the ASPIC has. This reflects a boundary in the model we want to solve, as the heating material cannot have a heat conductivity larger than the equivalent value obtained from the photonic circuit. This power budget problem can be addressed in several ways, from using a low loss waveguide (large-core or ridge silicon waveguides, or silicon nitride (SiN) waveguides) for reducing the distance between nodes. \\ 
In fact, this latter one comes with the need to assess the optical power out from every single node. As shown in figure \ref{Fig:fig4}B, most of the space is occupied by the grating couples and directional couplers from each node's output ports. This design has several limitations, such as footprint occupation and power budget, since it requires 90:10 couplers and low cross-talking cross-nodes. Using integrated photodetectors could improve the overall KPN footprint. Solutions, such as CLIPP\cite{morichetti2014non, carminati2017design}, could further enhance this design, as they have the feature of being transparent, and they can be directly integrated on the waveguide to be monitored. \\
A third element limiting the current design comes from the PCM as a tunable attenuator. In our case, the GSSe material has an absorption coefficient varying from 0.07 dB/$\mu$m to ~ 0.2 dB/$\mu$m by applying variable voltage pulses. However, having a single strip does not allow sufficient variable attenuation, limiting our capabilities to match the materials' proprieties in the heat distribution model. A solution that implements four strips of GSSe with different lengths between two KPNs can achieve much larger dynamic range in terms of loss modulation and further improve our system's accuracy. Integrating other materials could open the circuit to more configuration options, for example, implementing a proper "heat insulation" or "heat reflecting shield" by increasing the ER or producing a variation of the effective index that will return a backpropagating wave.\\ \\
The last point focuses on the boundary conditions. In our circuit every port towards the outside of the matrix are directly connected to a grating coupler for coupling all the light out. This can be seen in the heat model as a heat flux toward the outside. However, to fully generalize our ASPIC to all the possible model, we have to permit different configurations at each external port, for example switching between a grating coupler, that extracts all the optical signal, and a Sagnac Loop\cite{fernandez2019universal}, whose function is to reflect completely the light back into the matrix. This later solution could be used as heat insulation, providing an ASPIC that can solve more complex PDEs models.

In summary, we design and test an Application-specific Photonic Integrated Circuit using KPN capable of solving the analog partial differential equation for a heat distribution problem, with over 90\% accuracy. The Silicon Photonics integration permits to obtain a chip with a size smaller than a fingernail, over 50 nm optical bandwidth, and most importantly, obtain the results in a nanosecond scale. We show a passive and active version of the ASPIC, showing the higher accuracy we can achieve once we can tune the Phase Change Material between two nodes to match the properties of the model materials. With the progressive Through Silicon Via (TSV) and multi-layer technology\cite{zhang2020scalable}, we will be able to design the optical and electrical routes through different layers, achieving more complicated functionalities. Also, by add the ring modulators to the matrix, it is possible to reconfigure each node to achieve different splitting ratios, so that different types of circuit can be configure, from optical filter, multiplexer, de-multiplexer, and more complicated version for artificial intelligence and neuromorphic computing\cite{shastri2021photonics}.

\section{Method}

\paragraph{Device fabrication}
$ $ \\
The passive silicon photonic circuits are fabricated using an electron-beam lithography (EBL) process on a silicon on insulator (SOI) wafer by Applied Nanotools Inc. The complete circuit was designed using Klayout, a design framework for integrated circuitry. The GSSe deposition process is done by single-source thermal evaporation following previously disclosed protocols.\cite{zhang2019broadband} The microheater fabrication process is accomplished in two steps in the nanofabrication and imaging center at George Washington University. A tungsten titanium layer with $350 nm$ thickness represents the heater, route, and contact pads are evaporated using a DC sputter process. Then a $200 nm$ thick Aluminum layer is deposited on top of the tungsten titanium route and pads. (Figure 4)

\paragraph{Measurement setup}
$ $ \\
For the optical measurement, we used the Single-Die Probe Station as well as the measuring system from Maple Leaf Photonics.
For optical power transferred outside the ASPIC, we used two 8-parallel-port fiber arrays with 127 $\mu m$ pitch to inject the light in and capture the light out through grating couplers. The laser source and detector devices we use are Keysight N7778C and N7745A, respectively, both functionally programmed in the Maple Leaf Photonics measuring system. 
In the active measurement part, we use the National Instrument PXIe-5413 waveform generator module to generate the pulse for heating the PCM strips.
\paragraph{Heat transfer simulation}
$ $ \\
The heat transfer simulation mapping with the partial differential equation is performed using a two-dimensional finite element model in COMSOL Multiphysics. We use the stationary “heat transfer in solid” module, which elucidates the thermal boundary conditions. The simulated area is 5m x 5m, equally divided into 25 (5x5) squares using line segments. The equations applied are:
\begin{equation}
    d_{z}\rho C_{p}u\cdot \nabla T + \nabla\cdot q = d_{z}Q+q_{0}+d_{z}Q_{ted}
\end{equation}
\begin{equation}
    q = -d_{z}k\nabla T
\end{equation}
Where $\rho$ represents the density of the material ($kg/m^{3}$), $Q$ is the heat source, $u$ is the velocity, $q$ is the heat flux and $k$ is the thermal conductivity ($W(m\cdot K)$). In our stationary case, $d_{z}$ is out-of-plane thickness, $C_{p}$ represents the specific heat capacity($J/(kg\cdot K)$), and $Q_{ted}$ is the thermoelastic damping. \\
The initial value of the whole domain is 273 K, and the heat source temperature is 393 K, while all the remaining edges are set to have a heat flux toward the outside set at 293 K to represent the heat sink of this model.

\paragraph{Optical simulation}
$ $ \\
We use Lumerical Interconnect simulation software to emulate the photonic circuit on an integrated chip. S-parameter elements with user-defined files are employed to represent the Kirchhoff Photonic Node. The continuous-wave laser is placed to generate the light at a 1550 nm wavelength. Different S-parameter elements with simple parameter numbers are exploited to mimic the optical absorption change tuning function of the PCM. We also set the splitting ratio of optical waveguide coupler elements to characterize directional couplers. Oscilloscope elements are used to read and record the power output.

\section{Acknowledgement}

This work was performed in part at the George Washington University Nanofabrication and Imaging Center (GWNIC), as well as at the Harvard Center for Nanoscale Systems (CNS).
C.P. and J.H acknowledge funding support provided by NSF under award number 2132929.

\section{Contribution}

V.J.S. and T.E. conceived the original ideas, acquired the funds, and supervised the project. C.S designed and fabricated the photonic integrated circuit. C.P prepared the phase change material. C.S and N.P developed the relevant theories and analyses for the project, measured the PIC, and discussed the result. C.S processed the data and plotted the figures. J.M and X.M initiate the measurement environment. All authors discussed the results, contributed to writing the manuscript and provided they feedback.

\bibliographystyle{Science}
\bibliography{ASPIC.bib}

\newpage


\clearpage

\end{document}